\documentclass[journal=asbcd6,manuscript=article]{achemso}

\usepackage[version=3]{mhchem} 

\usepackage{eurosym}
\usepackage{mathtools} 
\usepackage{latexsym}
\usepackage{float}
\usepackage{color}
\usepackage{url}
\usepackage{graphicx}
\usepackage{color}
\usepackage{fancyhdr}
\usepackage{amscd}
\usepackage{booktabs}

\usepackage{algorithm}
\usepackage{algorithmic}

\usepackage{ulem}
\usepackage{chemarrow}
\usepackage{verbatim}

\usepackage{hyperref}

\newcommand{\junk}[1]{}

\definecolor{light-gray}{gray}{0.4}

\author{Andriy Didovyk}
\affiliation{BioCircuits Institute, University of California San Diego, La Jolla CA, USA}
\author{Oleg I. Kanakov}
\affiliation{Department of Radiophysics, Lobachevsky State University of Nizhniy Novgorod, Nizhniy Novgorod, Russia}
\author{Mikhail V. Ivanchenko}
\affiliation{Department for Bioinformatics, Lobachevsky State University of Nizhniy Novgorod, Nizhniy Novgorod, Russia}
\author{Jeff Hasty}
\affiliation{BioCircuits Institute, University of California San Diego, La Jolla CA, USA}
\alsoaffiliation{Department of Bioengineering, University of California San Diego, La Jolla CA, USA}
\alsoaffiliation{Molecular Biology Section, Division of Biological Science, University of California San Diego, La Jolla CA, USA}
\author{Ram{\'{o}}n Huerta}
\affiliation{BioCircuits Institute, University of California San Diego, La Jolla CA, USA}
\email{rhuerta@ucsd.edu}
\author{Lev Tsimring}
\affiliation{BioCircuits Institute, University of California San Diego, La Jolla CA, USA}
\email{ltsimring@ucsd.edu}

\title{Distributed classifier based on genetically engineered bacterial cell cultures}


\begin{document}

\begin{tocentry}
  \includegraphics{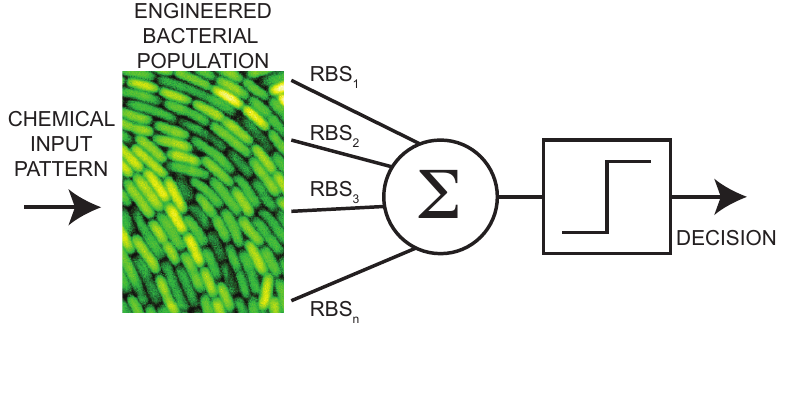}
\end{tocentry}

\begin{abstract}
We describe a conceptual design of a distributed classifier formed by a population of genetically engineered microbial cells. The central idea is to create a complex classifier from a population of weak or simple classifiers. 
We create a master population of cells with randomized synthetic biosensor circuits that have a broad range of sensitivities towards chemical signals of interest that form the input vectors subject to classification. The randomized sensitivities are achieved by constructing a library of synthetic gene circuits with randomized control sequences (e.g. ribosome-binding sites) in the front element. The training procedure consists in re-shaping of the master population in such a way that it collectively responds to the ``positive'' patterns of input signals by producing above-threshold  output (e.g. fluorescent signal), and below-threshold output in case of the ``negative'' patterns. The population re-shaping is achieved by presenting sequential examples and pruning the population using either graded selection/counterselection or by fluorescence-activated cell sorting (FACS). We demonstrate the feasibility of experimental implementation of such system computationally using a realistic model of the synthetic sensing gene circuits. 
\end{abstract}

\section{Introduction}

Pattern recognition and classification is the central topic in the field of machine learning~\cite{citeulike:873540}. Its applications span across disciplines such as computer vision~\cite{viola2001rapid}, natural language processing~\cite{sebastiani2002machine}, search engines~\cite{manning2008introduction}, medical diagnosis~\cite{kononenko2001machine}, classification of DNA sequences~\cite{goodman1998toward}, speech recognition~\cite{rabiner1993fundamentals},  computational finance~\cite{huertaelkan2013}, fraud detection~\cite{chan1998toward}, and many others~\cite{witten2011data}. In these contexts usually a system that solves a pattern recognition problem learns from the ``training'' data presented to it. Using the ``training'' data such a system forms an internal model to classify new previously unseen data. Typically these models are built in regular computers, although alternative approaches have been proposed~\cite{maass2002real,jaeger2004harnessing}. 

Many pattern recognition algorithms are biologically motivated. Biological organisms perform decision making based on classification of external environmental cues at all levels from intracellular (e.g. ref.~\citenum{ptashne1986genetic}) to organismal \cite{rangel2008framework} and even population-wide \cite{couzin2009collective}.  The development of artificial neural network (ANN) classifiers was inspired by the brain's natural ability to perform complex computational and classification tasks \cite{amit1992modeling}. The main principles of brain dynamics, its layered organization and ability to learn by adapting strengths of inter-neuron synaptic connections (plasticity) is mimicked in ANN by multilayered perceptrons and various learning algorithms~\cite{haykin1994neural}. Another learning approach is motivated by the adaptive immune system of jawed vertebrates which employs a population of immune cells with a diverse genetically encoded library of recognition specificities in order to implement learning, memory, and pattern recognition capabilities. The main principles of the natural adaptive immune system, its distributed nature and the ability to learn by positive and negative selection, have inspired the field of Artificial Immune Systems (AIS)~\cite{farmer1986immune,bersini1991hints,dasgupta1999overview}.

In this work we explore an alternative concept of building a classifier. Instead of using biological principles to motivate the design of computer-based classification systems, we propose to use synthetic biology to adapt biological systems themselves for solving complex classification tasks. The basic idea is to engineer populations of microorganisms capable of performing classification tasks based on consensus strategy. The desired binary classifier should produce a positive response (for example, above-threshold population-averaged fluorescence level) to positive inputs, and negative response (below-threshold fluorescence) to negative ones.  The learning algorithm then should consist of shaping the population in such a way that the population collectively ``arrives'' at the correct decision. 

The idea of aggregating many simple classifiers to yield a better classifier is a widely used strategy in machine learning. It is typically referred to as boosting \cite{schapire1990strength,freund1999short} and has evolved to inspire a variety of powerful algorithms used for pattern recognition like \mbox{AdaBoost} \cite{freund1995desicion} and bagging \cite{raey}. Boosting capitalizes on the idea that using a set of classifiers that produce barely better results than random guessing can achieve arbitrarily high accuracy when combined appropriately. Inspired by this idea, we propose to use genetically engineered cells with limited abilities to perform complex classification tasks. Given the groundwork in developing effective boosting algorithms, one can potentially build good classifiers made of such cells. 

Here we present a specific implementation of this general distributed pattern recognition system concept  using synthetic gene regulatory circuits in engineered cell populations. The proposed implementation requires to build a cell library with genetically encoded randomized sensitivities to external chemical signals to be classified. Such libraries have been successfully constructed for optimization of synthetic biological circuits~\cite{pfleger2006combinatorial,wang2009programming,zelcbuch2013}. This library forms a {\it master classifier} population that is pruned to learn how to solve a certain classification task. The learning is done by examples: cells with erroneous outputs are gradually attenuated from the population, while the ``correct'' cells are amplified. As a result of multiple iterations of pruning/amplification, a distributed classifier trained for a specific task emerges from the master population. We envision that an arbitrary external input subject to recognition can be encoded by a combination of chemical inputs capable of generating the engineered cellular response. In this paper, however, we will consider the most straightforward case when the vector of chemical concentrations {\it is} the input signal subject to classification. In the following we demonstrate the general principle of this classification procedure and describe its implementation using a model of a synthetic genetic circuit based on the lambda phage $P_{RM}$ promoter~\cite{ptashne1986genetic}. 

\section{Results and discussion}
 
\subsection{Learning by examples}
In this section we describe the general idea behind the training of a distributed classifier by presenting a set of positive and negative examples. In the following, we denote the set of input variables to be classified as $\mathbf{x}\in \Re^M$. A classifier is the function $y = 2H\left(f(\mathbf{x})-\theta\right)-1$, such that if $f(\mathbf{x})>\theta$, the answer is $y=1$, and otherwise, $y=-1$. Here $H(\cdot)$ is the Heaviside function, $\theta$ is a scalar threshold, and $f(\mathbf{x})$  is the scalar function of the inputs. The  heart of the pattern classifier is the function $f(\mathbf{x})$ that minimizes the  classification errors for a given distribution of positive and negative inputs. 

In general, the classifier function is not known {\it a priori} and has to be learned by training the classifier using examples (training data). 
The training of a classifier by examples consists in finding a function $f(\mathbf{x})$ that minimizes the error in mapping of a given set of $N$ examples $\mathbf{x}_i$ to a set of binary  outputs, $y_i=\{-1,+1\}$ which label the examples. In the following, we will say that if the output $y_i=+1$, the example $i$ belongs to the ``positive''  class and  if the output is $y_i=-1$ the example $i$ belongs to the opposing ``negative'' class. 

In our proposed population-based classifier, $\mathbf{x}$ will be a set of concentrations of chemical signals which cells are subjected to, and the cells are assumed to contain  gene circuits that produce a fluorescent signal $z_i(\mathbf{x})$ in response to  the input signal $\mathbf{x}$. The overall signal function, $f(\cdot)$,  will be the normalized linear sum of the individual fluorescence signals from all $N_c$ cells in the trained population:
\begin{equation}
f(\mathbf{x})=\frac{1}{N_c}\sum_{i=1}^{N_c}z_i(\mathbf{x}). \label{eq:principal}
\end{equation}
The key to the ``trainability'' of the distributed classifier is to prepare a master population of cells with broadly varied functions $z_i(\mathbf{x})$, so this population can be appropriately shaped to perform the needed classification task. This can be achieved using synthetic gene regulatory circuits with randomized control elements such as promoter regions, ribosome binding sites, or other sequences as described in detail below. 

We assume that the master population contains cells that individually provide correct answers to subsets of the data to be classified, but that in general no single cell provides correct answers to all data (weak classifiers). The goal of training is therefore to shape the master population to create a distributed consensus classifier that performs better then any cell individually. Such training must amplify the cells providing correct answers as frequently as possible and conversely suppress the cells with poor overall performance. Thus, unlike the typical machine learning procedure that modifies the parameters of the system, our proposed learning by examples consists in modifying the composition of the cell population based on the examples with known outcomes. The details of our training algorithm for the specific implementation of a genetic sensing circuit are outlined below.

\subsection{Distributed classification with randomized synthetic gene sensors}

\begin{figure}[ht!]
  \centering
  \includegraphics[width=0.75\linewidth]{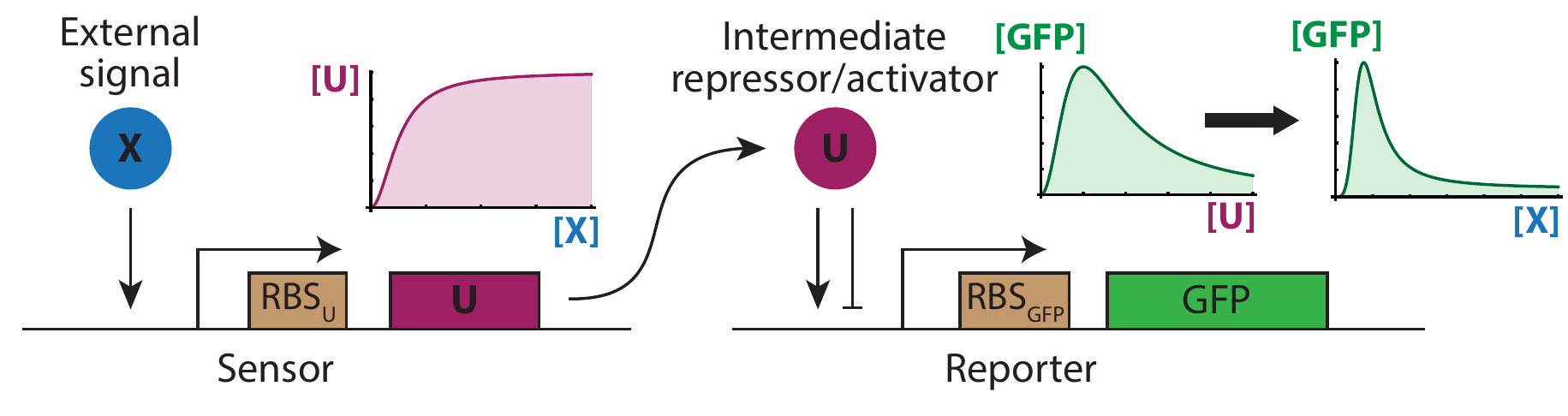}
 
  \caption{A modular genetic circuit proposed for implementing a distributed genetic classifier. Sensing and response functionalities are split into separate modules. In the first module (sensor), an inducible promoter drives the expression of the transcription factor $U$ in response to the applied signaling molecule $X$. The response function of the promoter is chosen to be monotonic (see inset). In the second module (reporter), another inducible promoter drives the expression of a reporter (GFP) in response to induction by $U$. The promoter is activated by intermediate concentrations of $U$ and inhibited by high concentrations of $U$. Thus the resulting response function of the entire two-promoter circuit to the concentration of signaling molecule is bell-shaped for the relevant values of the circuit parameters as shown on the inset.}
  \label{fig1ad}
\end{figure}

Here we outline an implementation of the proposed distributed classifier with a diverse population of bacterial cells containing synthetic sensory genetic circuits with randomized parameters. For simplicity we focus on a scalar input with only one chemical signal affecting the gene circuit, however the same approach can be straightforwardly generalized to the multi-dimensional vector input. In our two-gene design (Fig.~\ref{fig1ad}) the sensing and the reporting functionalities are split between the two genetic modules. The sensing module is monotonically induced by the external chemical signal $X$ and drives the synthesis of a transcription factor $U$. The second promoter is regulated by $U$ and drives the expression of a reporter protein, for example, green fluorescent protein (GFP). The reporter promoter is activated by $U$ at intermediate concentrations and inhibited at higher concentrations, thus being active only within a finite range of concentrations of $U$. The classic well-characterized example of such promoter is the promoter $P_{RM}$ of phage lambda which is activated by intermediate concentrations and is repressed by high concentrations of the lambda repressor protein CI~\cite{ptashne1986genetic}. 

This two-gene circuit can be modeled by the following set of biochemical reactions
\begin{align*}
\emptyset&\autorightarrow{$r_u(x)$}{}U;\quad
U\autorightarrow{$\mu_u$}{}\emptyset \\
\emptyset&\autorightarrow{$r_g(u)$}{}GFP; \quad
GFP\autorightarrow{$\mu_{g}$}{}\emptyset
\end{align*}
where $x$ and $u$ are the concentrations of $X$ and $U$, $r_u(x)$ and $r_g(u)$ are the effective production rates of $U$ and $GFP$, respectively, and $\mu_u$ and $\mu_g$ are the degradation rates of $U$ and $GFP$. The rates of gene expression in this system can be described by standard Hill functions:
\begin{eqnarray}
r_u(x;m_u) = m_u \cdot \frac{\alpha{}A_u^{p_u}+x^{p_u}}{A_u^{p_u}+x^{p_u}}\\
r_g(u;m_g) = m_g \cdot \frac{A_g^{p_g} u^{p_g}}{(A_g^{p_g}+u^{p_g})^2}
\end{eqnarray}
where $\alpha$ describes the basal expression from the sensor promoter in the absence of the signaling molecule $X$, $A_u$ is the  dissociation constant of $X$ with the sensor promoter, the Hill coefficient $p_u$ characterizes the cooperativity of activation of the sensor promoter, $p_g$ characterizes the cooperativity of activation and repression of the reporter promoter by the transcription factor $U$, $A_g$ is the dissociation constant for activation and repression of the reporter promoter by $U$ (we assume the activation and repression cooperativities and dissociation constants to be the same), $m_u$ and $m_g$ describe the overall strength of production of $U$ and GFP respectively.  

In mass-action approximation the dynamics of GFP production in this system is described by the following system of ordinary differential equations:
\begin{eqnarray}
		\frac{d\,u}{dt}&=& r_u(x;m_u) - \mu_{u} u \label{model-eq1}\\
		\frac{d\,z}{dt}&=& r_g(u;m_g) - \mu_{g} z
		\label{model-eq2}
\end{eqnarray}
where $z$ is the concentration of GFP. The steady state concentration of GFP as a function of the concentration of the external chemical signal $X$ can be easily found from the equations (\ref{model-eq1}) and (\ref{model-eq2}) :

\begin{equation}
z^*(x;m_u,m_g) = \frac{r_g(r_u(x;m_u)/\mu_u;m_g) }{ \mu_g } \\
\label{eq:1D_transf_functn}
\end{equation}

\begin{figure}[ht]
  \centering
  \includegraphics[width=0.6\linewidth]{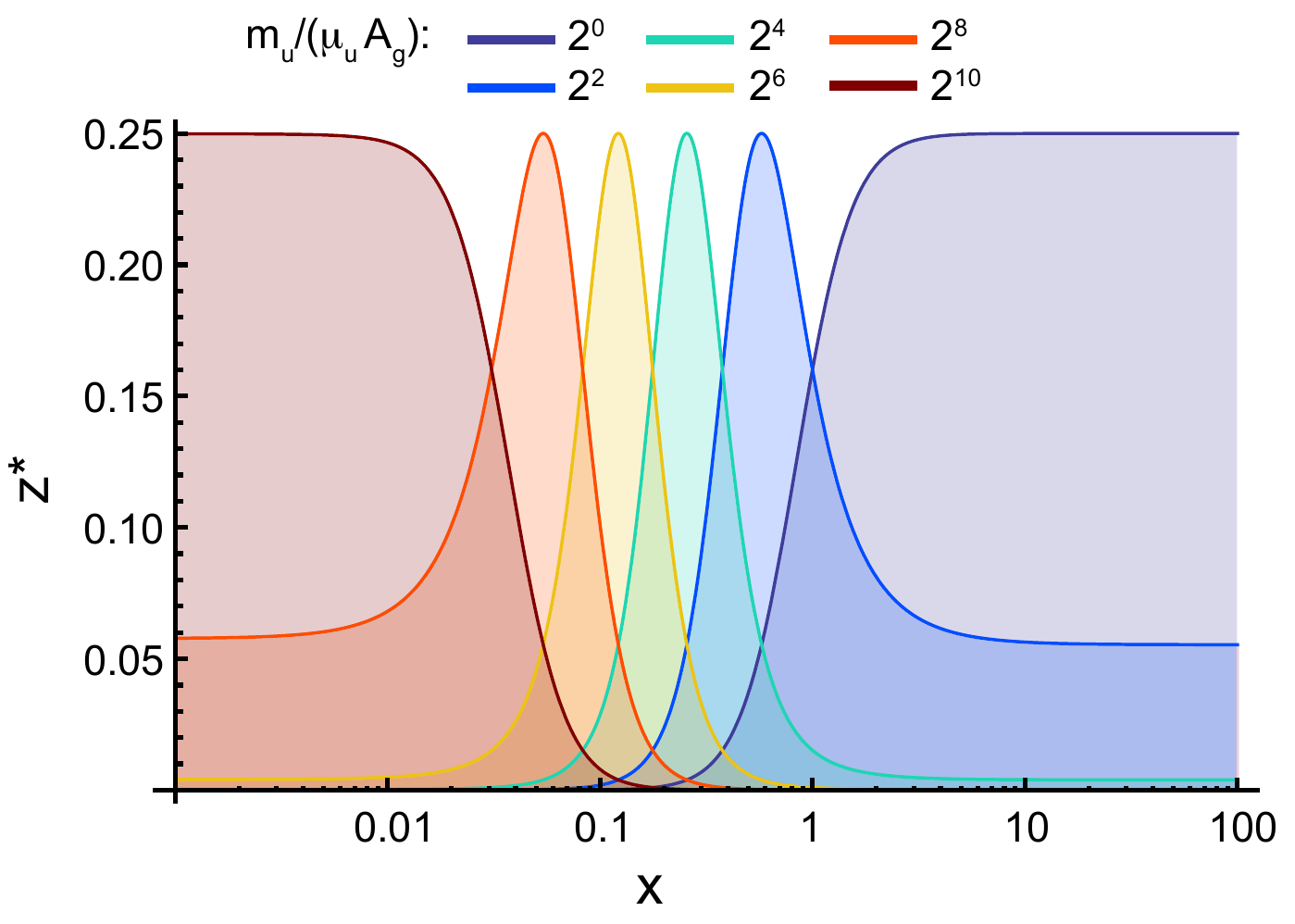}
 
  \caption{Steady state GFP concentration ($z^*$) as a function of the concentration of the external chemical signal $X$ for the modular classifier circuit shown for a range of $m_u$ values representing a range of the relative strengths of the sensor promoter (Fig.~\ref{fig1ad}). Non-dimensional circuit parameters are $\mu_u=\mu_g=m_g=A_u=1$, $A_g=20$, $p_g=p_u=2$, $\alpha=10^{-3}$. }
  \label{fig2ad}
\end{figure}

 Function $z^*(x)$ is bell-shaped in a broad range of $m_u/\mu_u \in (A_g, A_g/\alpha)$~(Fig.~\ref{fig2ad}). Varying $m_u/\mu_u$ allows to create a library of circuits which act as low-pass filters ($m_u/\mu_u \le A_g$), high-pass filters ($m_u/\mu_u \ge A_g/\alpha$), or tunable bandpass filters for the intermediate values of $m_u/\mu_u$. As described in the following sections, such a library can be used to train a cell population-based distributed classifier. 
Since common sensor promoters can have regulatory range of over $10^3$ ($\alpha=10^{-3}$)~\cite{lutz1997independent, cox2007programming} in order to create a library that at the same time contains low-pass, bandpass, and high-pass filters $m_u/\mu_u$ ratio has to be varied at least $1/\alpha = 10^3$ fold within $[A_g, A_g/\alpha]$ range~(see Fig.~\ref{fig2ad}). 
Such libraries have been widely constructed experimentally: $m_u$ can be varied over more than $10^5$ fold range by varying the DNA sequence within and near the ribosome binding site of the gene of interest~\cite{salis2009automated, kudla2009coding};  this range can be further expanded by modulating the sensor promoter strength~\cite{brewster2012tuning}; $m_u/\mu_u$ ratio can also be modulated by varying the stability of the $U$ coding mRNA as well as the stability of the protein $U$ itself~\cite{carrier1999library, wang2008tuning, flynn2001overlapping, pfleger2006combinatorial}. 

The modular architecture of the classifier circuit proposed here allows one to independently select and optimize the sensor and the reporter functionalities in order to maximize the classifier performance. Well-characterized sensor promoters that can be induced by a variety of chemical signals with the appropriate monotonic response are common~\cite{lutz1997independent, zucca2012characterization, cox2007programming, lee2005propionate, khlebnikov2001homogeneous} and can be readily combined with the reporter promoter such as the well characterized lambda phage promoter $P_{RM}$~\cite{ptashne1986genetic, isaacs2003prediction}. For these reasons the proposed two-gene sensory circuit appears to be well-suited for experimental implementation of a distributed cell population based classifier.


\subsection{Classifier training algorithm}
In order to train a classifier we need to be able to sort individual cells based on their response to a set of known examples. A hard-decision algorithm would imply that if a given example $x_i$ belongs to a positive class, $y_i=+1$, and the GFP level in $j$-th cell is above a threshold $z_j^*\left(x_i\right)>\theta$, then that particular cell should be retained in the population since the cell is providing the correct answer. Meanwhile, the cells not reaching the threshold level of expression after presenting a positive example should be removed from the population. On the other hand, if a negative example is presented ($y_i=-1$), the cells generating above-threshold fluorescence should be eliminated and the cells which are below threshold should be retained. By this selection mechanism we would ensure that only the cells that generate correct answers will survive.  

However, this hard-decision training algorithm in most practical situations would lead to poor performance. As mentioned above, in general each cell is a weak classifier and so it can not provide the complete classification solution. If positive and/or negative categories encompass a broad range or several distinct ranges of inputs, a sub-population in which all cells would generate above-threshold output for positive examples and below-threshold output for negative examples would be empty. Thus, an outright elimination of all cells producing ``incorrect answer'' to any of many training examples would eventually lead to elimination of all cells.  To avoid this undesirable outcome, a ``soft-training'' algorithm has to be employed in which even the ``erroneous'' cells have a chance to remain in the population, and so the resultant population-based classifier will produce the correct answer only by the population average, not the unanimous decision of all cells.

\begin{figure}[ht]
  \centering
  \includegraphics[width=0.8\linewidth]{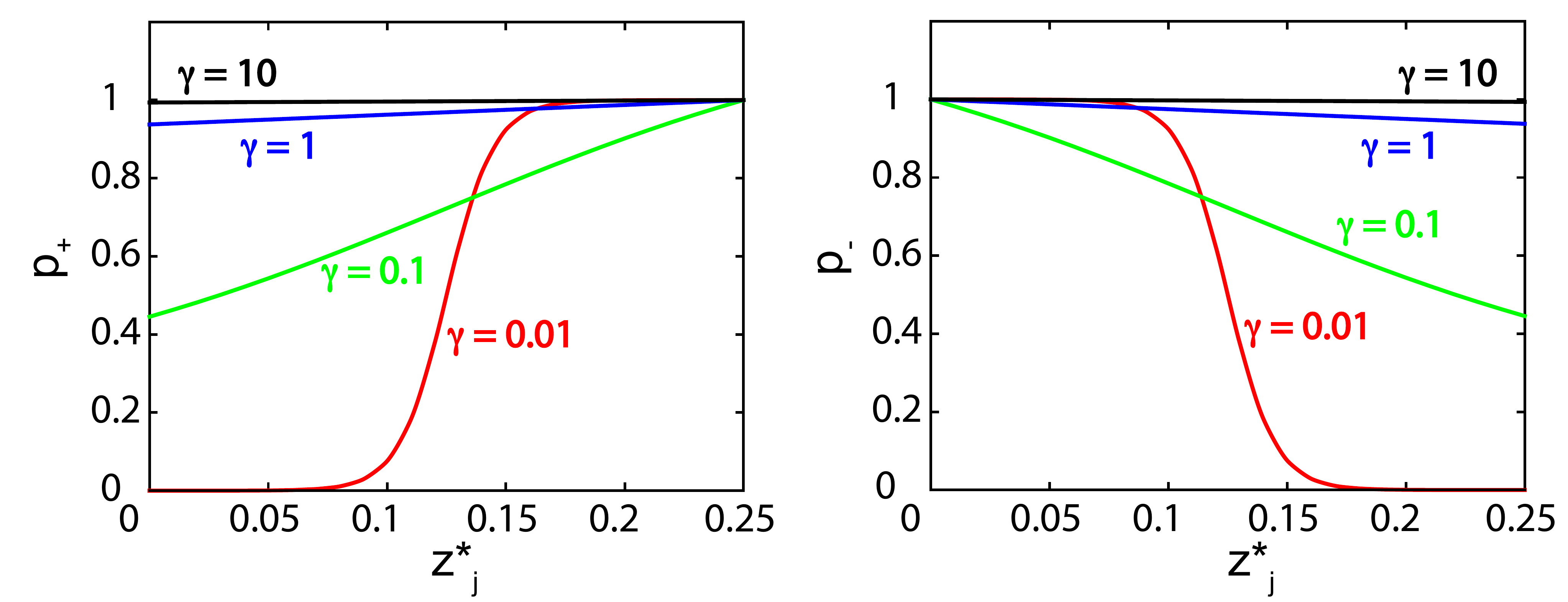}
 
  \caption{Cell survival probabilities during training upon presentation of a positive ($p_{+}(z_j^*; \gamma)$) or a negative example  ($p_{-}(z_j^*; \gamma)$). }
  \label{fig_p_vs_gamma}
\end{figure}

We postulate that the elimination of cells from the population is governed by two sigmoidal cell survival probability functions $p_{+}(z_j^*) = (1 + \xi)^{-1} + (1 + \xi \exp(-z_j^* / \gamma ))^{-1}$  and  $p_{-}(z_j^*) = 1+(1 + \xi)^{-1} - (1 + \xi \exp(-z_j^* / \gamma ))^{-1}$ for positive and negative examples correspondingly, where $\xi = \exp((8 \gamma)^{-1})$ and  $0 \le z_j^* \le 1/4$ are chosen to encompass the entire range of possible values of $z_j$ (eq.(\ref{eq:1D_transf_functn})) (Fig.~\ref{fig_p_vs_gamma}).  These functions are chosen such that the cells that respond perfectly to the presented example ($z_j^*=1/4$ or $z_j^*=0$ for a positive or a negative example correspondingly) are retained in the population ($p_{\pm}(z_j^*)=1$). Otherwise the cells are eliminated from the population with the probability that depends on the cell fluorescence $z_j^*$ and the ``rigidity'' of training $\gamma$. For very small $\gamma$ the cells not responding to the currently presented input are eliminated with high probability. This can potentially eliminate many cells that are required to recognize other positive examples $x_i$ from the pattern being taught leading to the poor overall performance. For too large $\gamma$ (weak elimination), the learning rate slows down significantly requiring unreasonably large number of training iterations in order to achieve high performance.  Therefore an optimal value of the parameter $\gamma$ can be chosen to balance performance with the learning speed. In general this value will be different for each individual problem. 
As usual in the design of classifiers, the output element of the classifier must be a threshold element. In the distributed classifier described here, the mean population fluorescence $f(x)$ has to be compared with a suitably chosen threshold $\theta$ such that the above-threshold fluorescence ($f(x)>\theta$) corresponds to the positive class, and below-threshold fluorescence ($f(x)<\theta$) to the negative class. The optimal threshold $\theta$ can be found by presenting the training set of examples and minimizing the percentage of incorrect answers. The training procedure described above can be formalized as Algorithm~\ref{table:CV}. 

\begin{algorithm}[h!]
\caption{Algorithm for training the gene expression classifier.}\label{table:CV}
\begin{algorithmic}
\STATE {\bfseries Input:} The data set $x_i$ with $N$ examples. The class type $y_i$ for each example. The total number of cells $N_c$. The number of iterations of the algorithm, $N_{iter}$.
\STATE {\bfseries Output:} The trained set consisting of $N_c$ selected cells; classification threshold $\theta$.
\FOR{iteration $k \leftarrow 1$ to $N_{iter}$}
   \STATE Choose a random  example from $x_i$ in the range $i=1,\ldots,N$ 
   \COMMENT{Apply the chosen chemical input to the cells.}
   \FOR{$j \leftarrow 1$ to $N_c$}
	   \STATE{Eliminate the cell with probability $1 - p_{\pm}(z_j^*(x_k))$ (for $y_i = \pm 1$) with random cell replacement from the previous iteration or from the master population for $k=1$. }
   \ENDFOR
\ENDFOR\\
\FOR{training example $i \leftarrow 1$ to $N$}
\STATE{Use selected cells to measure the mean population fluorescence:}
\STATE $f(x_i) \leftarrow N_c^{-1}\sum_{j=1}^{N_c} z_j(x_i)$
   \ENDFOR
   \\
{\bf Find} threshold $\theta$ for which $\sum_{k=1}^Ny_k [2H(f(x_k)-\theta)-1]$ is maximal
\end{algorithmic}
\end{algorithm}



In the following sections we use this algorithm to train a computational model of  the distributed gene expression classifier and analyze its performance using sets of simulated data. We generate these sets using the model of the synthetic gene circuit (\ref{eq:1D_transf_functn}) presented above. In order to simulate the combined effects of biological and instrumental noise on the performance of the classifier we will assume that the resultant mean fluorescence of the population contains both additive and multiplicative noise:
\begin{equation}
f(x)=\frac{1}{N_c}\sum_{i=1}^{N_c}z_i(x)(1 + \varepsilon) + \zeta. 
\label{eq:principal_w_noise}
\end{equation}
where $\varepsilon$ and $\zeta$  are independent normal random variables with the respective means of $0$ and $\sigma/4$ and standard deviations $\sigma$ and $\sigma/4$.
As usual in performance evaluation, each data set is divided into two parts, one for training and another for testing, and the overall performance of the classifier is measured as the percentage of correct answers on the test sets. The classifier performance is calculated using Algorithm~\ref{table:CV2}.

\begin{algorithm}[h]
\caption{Algorithm for testing the gene expression classifier.}\label{table:CV2}
\begin{algorithmic}
\STATE {\bfseries Input:} The testing data set $x_i$ with $N$ examples. The class type $y_i$ for each example. The total number of cells $N_c$.  
\STATE {\bfseries Output:} The performance of the classifier.
\STATE{Use all the cells selected during training.}
\STATE{correct $\leftarrow$ 0}
\FOR{$i\leftarrow 1$ to $N$}
\STATE{Measure average population fluorescence $f(x_i)$} 
\IF{$f(x_i)>\theta$ \AND $y_i=1$}
    \STATE{correct$=$correct$+1$}
\ENDIF
\IF{$f(x_i)<\theta$ \AND $y_i=-1$}
    \STATE{correct$=$correct$+1$}
\ENDIF
\ENDFOR
\STATE{performance$=$correct$/N$}
\end{algorithmic}
\end{algorithm}

\begin{figure}[h!]
\begin{center}
\includegraphics[width=0.8\linewidth]{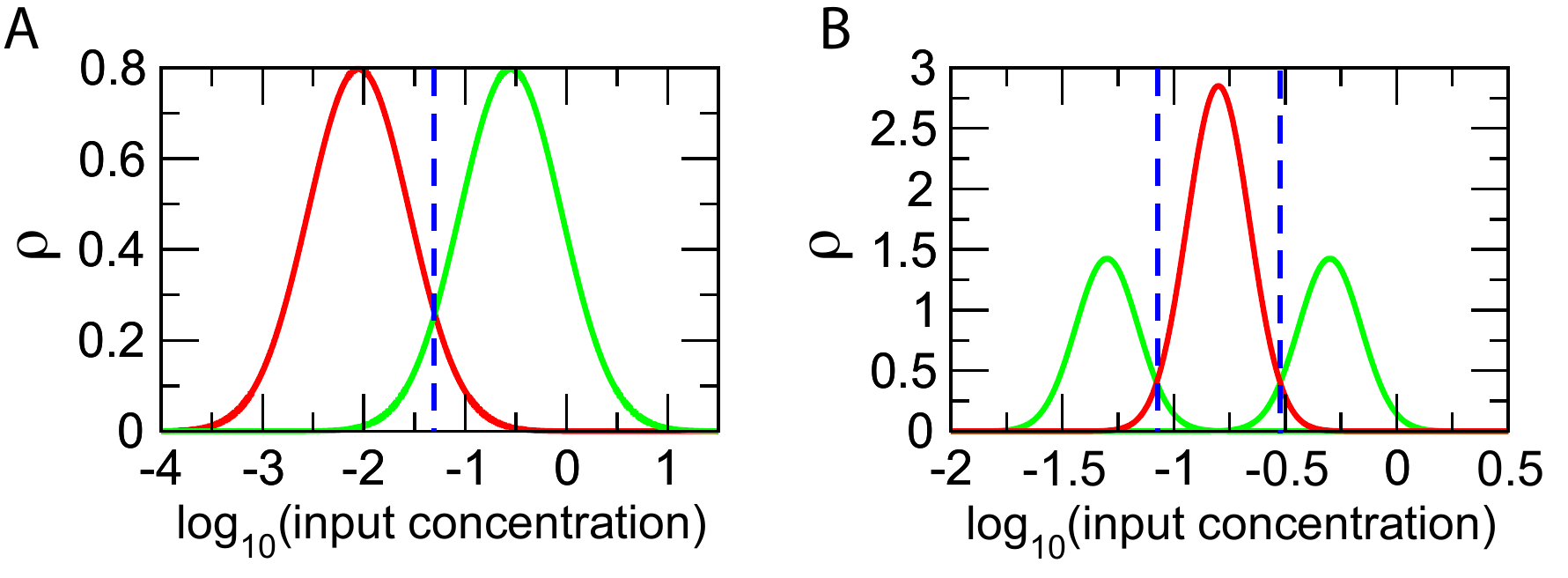}
\end{center}
 \caption{The data is generated from two (left) or three (right)  log-normal distributions generating the positive class (green) and the negative class (red) examples. The optimal thresholds for discriminating between the classes are represented by the dashed vertical lines. The maximum performance achievable by a classifier trained on infinite amount of examples from the two distributions is 93.3\%, from the three distributions is 94.8\%. The minimum in both cases is 50\% which is equivalent to a random answer selection. }
  \label{fig_theoretical performance}
\end{figure}

We will start with the case of discriminating data generated by two log-normal distributions of input chemical concentration with some overlap between the two classes~(Fig.~\ref{fig_theoretical performance}A). This is an example that illustrates that the distributed classifier can handle even non-separable problems. After that we test the classifier on a more challenging problem of discriminating the data generated by complex distributions (one unimodal and another bimodal), when the negative class is surrounded by the positive class on both sides~(Fig.~\ref{fig_theoretical performance}B).

\subsection{Discrimination of two log-normal classes in one dimension}

\begin{figure}[h!]
\begin{center}
\includegraphics[width=1.0\linewidth]{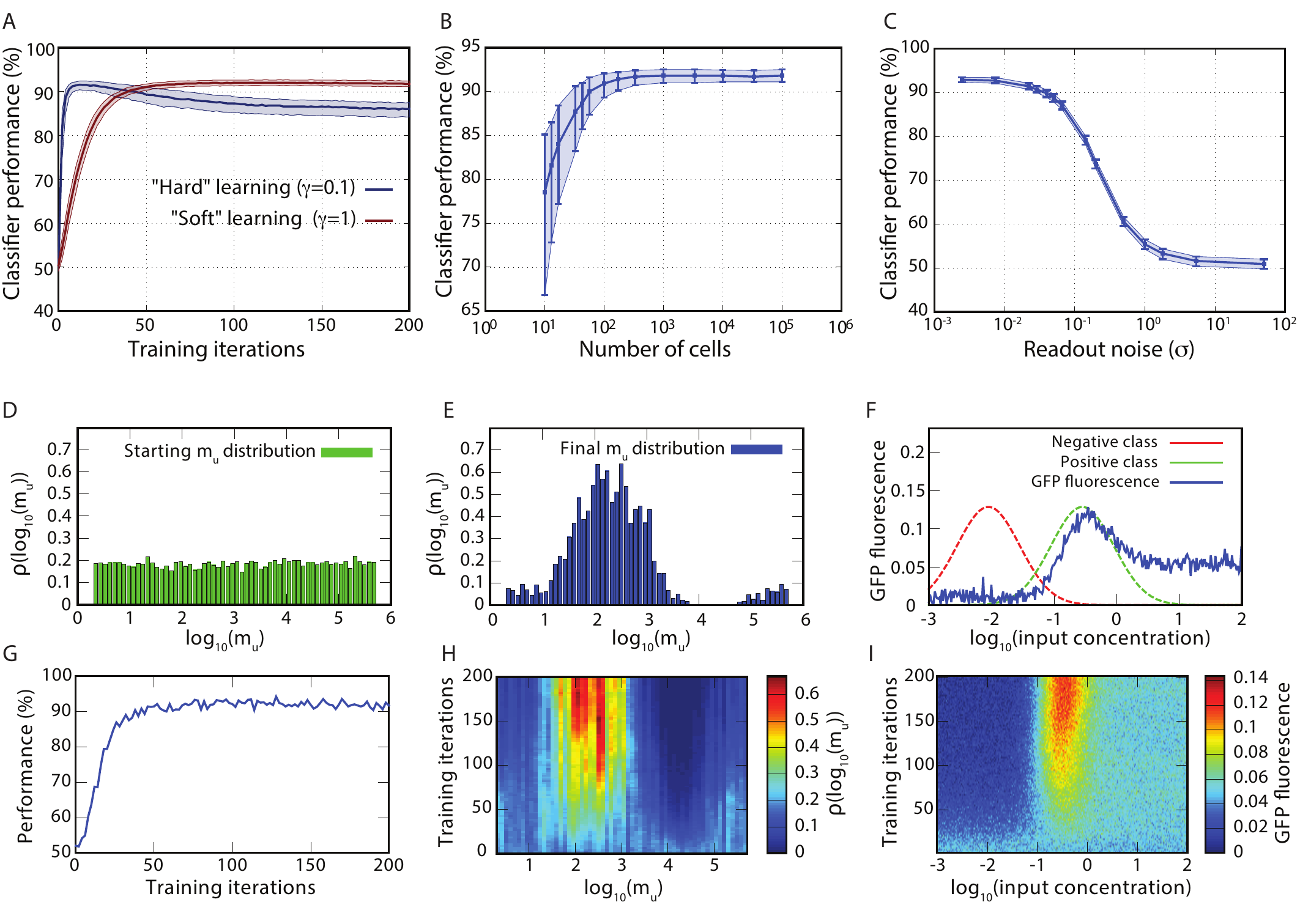}
\end{center}
 \caption{Classification results for the data set drawn from two log-normal distributions, Fig.~\protect\ref{fig_theoretical performance}A. (A) Evolution of the classifier performance for $\gamma=0.1$ (``hard'' learning; blue) and $\gamma=1$ (``soft'' learning; red), population size $N_c=10^4$ cells. The classifier performance versus cell population size $N_c$ (B) or GFP fluorescence readout noise $\sigma$ (C); $\gamma=1$ in (B) and (C), $N_c=10^4$ in (C). The median and interquartile range of the distribution of the classifier performance calculated from $10^3$ different stochastic realizations are shown in (A)--(C), readout noise $\sigma=1/35$ in (A) and (B). (D)--(I)  Evolution of the parameters of the classifier before and after training -- an example trajectory. The parameters used are $\gamma=1$, $N_c=10^4$, $\sigma=1/35$. It illustrates the shift in the distribution of parameters due to the training process of elimination of cells. The distribution of RBS/promoter strengths $m_u$ before training (D) and after 200 training iterations (E). (F) Normalized GFP fluorescence of the ensemble of cells $f(x)$ (blue) after 200 training iterations, log-normal distributions generating positive (green) and negative (red) class examples. (G) Evolution of the classifier performance in this realization. Evolution of $m_u$ distribution (H) and normalized cumulative GFP fluorescence $f(x)$ (I).}
  \label{fig2}
\end{figure}

As the first example we used the data drawn from two log-normal distributions centered at log(x)=-0.55 (class $+1$) and log(x)=-2.05 (class $-1$) with standard deviation of 0.5~(Fig.~\ref{fig_theoretical performance}A). The data has been generated on a log-scale since it is the natural scale of the response of the genetic circuit used in the classifier (Fig.~\ref{fig2ad}). Fig.~\ref{fig_theoretical performance}A gives an illustration of the two generating distributions.  The optimal theoretical performance is determined by choosing the threshold that separates the positive-class distribution from the negative one in a manner that minimizes discrimination errors. In this particular case, the optimal threshold value that leads to the minimal number of errors and therefore to the maximum performance is located exactly in the middle between the peaks of the two distributions which is indicated by the blue dashed vertical line ($\log(x)=-1.3$).  Due to the nonseparability of the two classes it is impossible to solve this classification problem with 100\% success rate. The optimal theoretical performance for this problem is 93.3\%. An ideal classifier could approach this value for infinite amount of training and evaluation data. 

Using a ``hard'' learning strategy (small $\gamma=0.1$), the population-based classifier achieves high performance of 91.6\% in just 12 iterations, however with further training the performance deteriorates achieving 86.1\% after 200 iterations (Fig.~\ref{fig2}A). These results are obtained using $N_c=10^4$ cells (all other circuit parameters are as described in Fig.~\ref{fig2ad}). Such deterioration of performance is a well known phenomenon in machine learning, where early stopping is often applied in similar situations~\cite{rosset2004boosting,zhang2005boosting}. 

In contrast, as discussed above, a ``soft'' learning strategy ($\gamma=1.0$) allows one to achieve the maximum performance (92.0\%) with high robustness over a wide range of training iterations (Fig.~\ref{fig2}A). Relatively small number of cells is sufficient to achieve such performance (Fig.~\ref{fig2}B). The classifier is also robust to readout noise (Fig.~\ref{fig2}C). 

The evolution of the parameters of the classifier during ``soft'' training is demonstrated in Fig.~\ref{fig2}D--I. The training process leads to a robust selection of the cells with those values of the genetic diversity parameter $m_u$ that allow the normalized population-wide GFP response $f(x)$ to roughly track the positive class data distribution~(Fig.~\ref{fig2}E--F, H--I). Correspondingly it is possible to reliably select such a threshold $\theta$ (Algorithm~2) that allows the classifier to achieve performance close to the theoretical maximum for this problem after roughly 50 training iterations~(Fig.~\ref{fig2}G).

\subsection{Discrimination of a bimodal vs a unimodal class in one dimension}

\begin{figure}[h!]
\begin{center}
\includegraphics[width=1.0\linewidth]{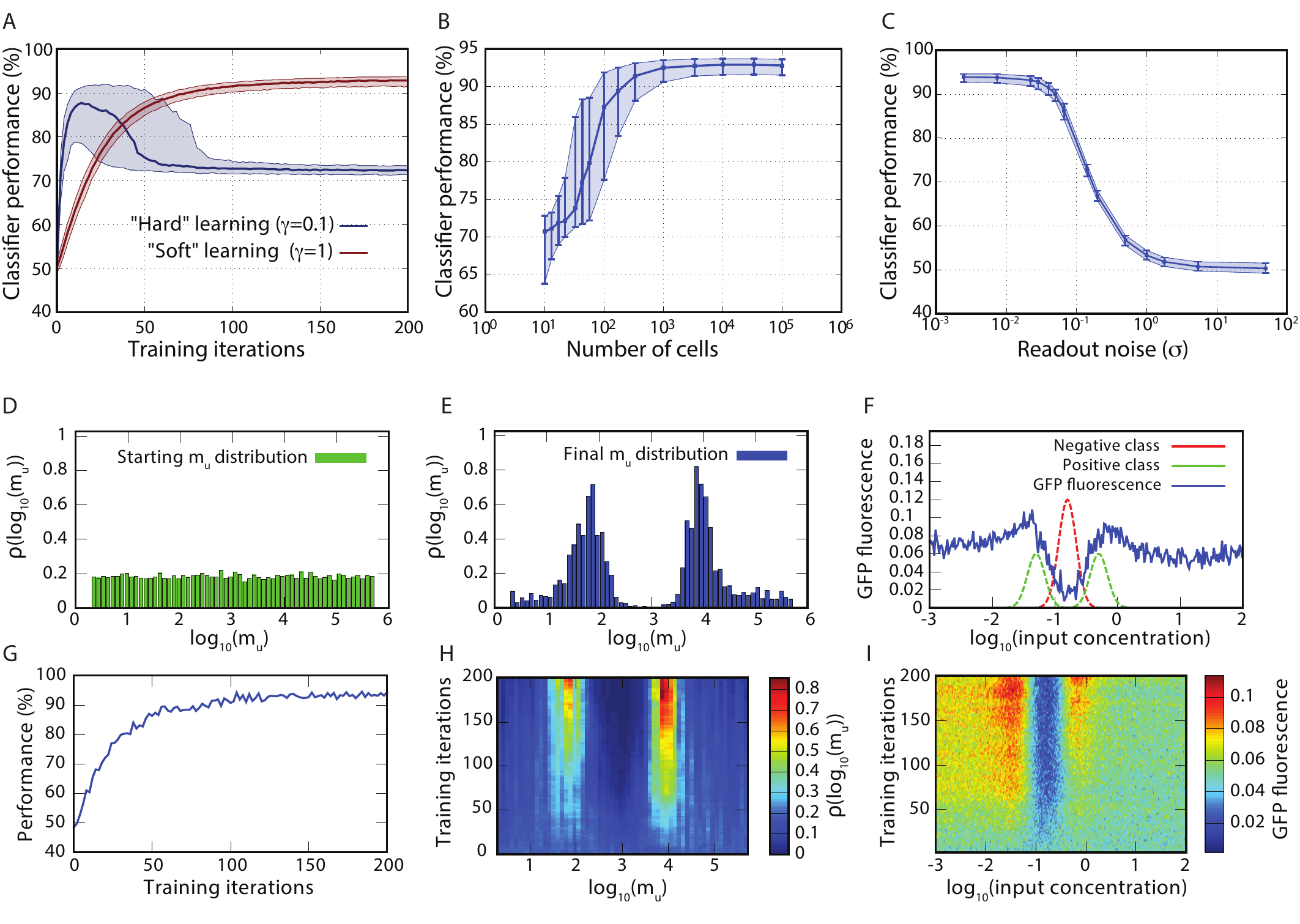}
\end{center}
 \caption{Classification results for the data set drawn from one bimodal and one unimodal distributions, Fig.~\protect\ref{fig_theoretical performance}B. (A) Evolution of the classifier performance for $\gamma=0.1$ (``hard'' learning; blue) and $\gamma=1$ (``soft'' learning; red), population size $N_c=10^4$ cells. The classifier performance versus cell population size $N_c$ (B) or GFP fluorescence readout noise $\sigma$ (C); $\gamma=1$ in (B) and (C), $N_c=10^4$ in (C). The median and interquartile range of the distribution of the classifier performance calculated from $10^3$ different stochastic realizations are shown in (A)--(C), readout noise $\sigma=1/35$ in (A) and (B). (D)--(I)  Evolution of the parameters of the ensemble of cells before and after training -- an example trajectory. The parameters used are $\gamma=1$, $N_c=10^4$, $\sigma=1/35$. It illustrates the shift in the distribution of parameters due to the training process of elimination of cells. The distribution of RBS/promoter strengths $m_u$ before training (D) and after 200 training iterations (E). (F) Normalized GFP fluorescence of the ensemble of cells $f(x)$ (blue) after 200 training iterations, log-normal distribution generating positive (green) and negative (red) class examples. (G) Evolution of the classifier performance in this realization. Evolution of $m_u$ distribution (H) and normalized cumulative GFP fluorescence $f(x)$ (I).}
  \label{fig2.3gaussians}
\end{figure}

This is a more challenging one-dimensional classification problem where a negative class is ``sandwiched'' by the positive class on both sides~(Fig.~\protect\ref{fig_theoretical performance}B).  The negative class data is generated by a log-normal distribution centered at log(x)=-0.8. The positive class data is generated from two equivalent log-normal distributions centered at log(x)=-1.3 and -0.3. The standard deviation of all three distributions is 0.14. The distributions are normalized such that on average the equal number of examples is drawn from the negative and positive classes. The maximum theoretical performance on this problem is 94.8\%, determined as in the example above. We use the same genetic circuit parameters as in the previous example (see Fig.~\ref{fig2ad}). 

This classification problem exemplifies the differences between the ``hard'' and ``soft'' learning strategies (Fig.~\ref{fig2.3gaussians}A). Similarly to the previous example the ``hard'' learning strategy ($\gamma=0.1$) initially leads to rapid improvement of classifier performance, reaching the maximum performance of 87.7\% in 14 iterations. However unlike the previous example this increase is not robust and is characterized by high stochastic variability~(compare Fig.~\ref{fig2}A and Fig.~\ref{fig2.3gaussians}A). With further training the performance quickly degrades due to stochastic extinction of the cells fitting one of the two positive class distribution peaks (see the details below). These problems can be ameliorated by employing ``soft'' training strategy ($\gamma=1$)~(Fig.~\ref{fig2.3gaussians}A). In this case the maximum performance is higher (92.9\%) and is achieved much more reliably albeit in noticeably higher number of training iterations. Similarly to the previous example, relatively low number of cells is sufficient to achieve this performance (Fig.~\ref{fig2.3gaussians}B). The performance is robust with respect to the readout noise~(Fig.~\ref{fig2.3gaussians}C). An example of the evolution of the parameters of the classifier during ``soft'' training is shown on Fig.~\ref{fig2.3gaussians}D--I.

\subsection{Analytical description of soft-learning performance}

\begin{figure}[h!]
\begin{center}
\includegraphics[width=1.0\linewidth]{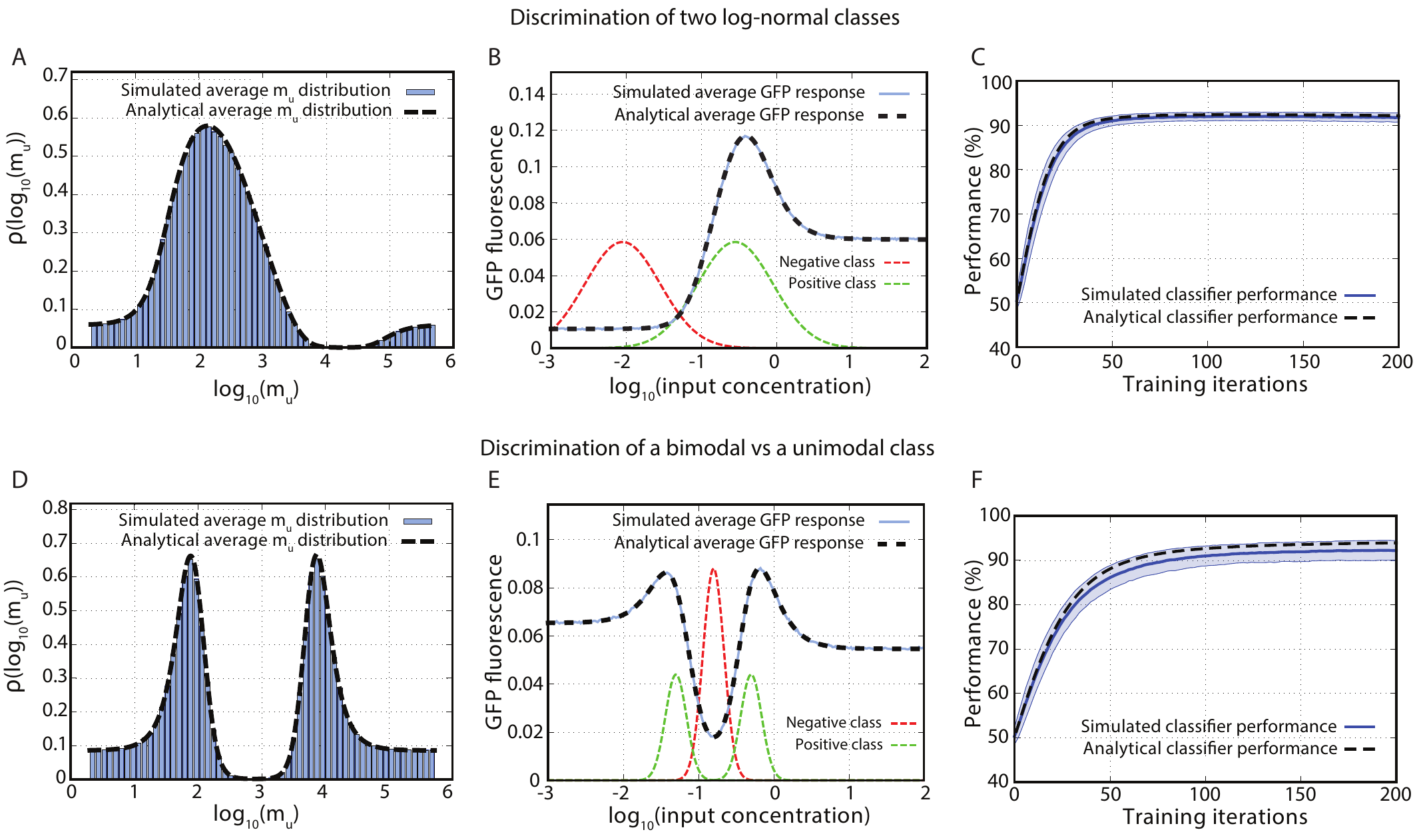}
\end{center}
 \caption{ Comparison of the analytical theory with the simulation results for the case of soft-learning ($\gamma=1$). (A)--(C) Discrimination of two log-normal classes. (D)--(F) Discrimination of a bimodal vs a unimodal class. (A), (D) Ensemble average distribution of the genetic diversity parameter $m_u$ calculated after 200 training iterations from $10^3$ independent simulations (blue bars) vs the analytical prediction (black dashed line). (B), (E) Ensemble average population-wide GFP responses calculated after 200 training iterations from $10^3$ independent simulations (blue line) vs the analytical prediction (black dashed line). (C), (F) Ensemble average classifier performance calculated from $10^3$ independent simulations vs the analytical prediction; the simulation results are shown as mean $\pm$ standard deviation. All other simulation parameters are the same as in Fig.~\ref{fig2}D--I,~\ref{fig2.3gaussians}D--I}
  \label{fig.analytics}
\end{figure}

We have developed an analytical theory that describes the training and the resulting performance of the cell-based classifier in the limit of ``soft'' learning (see Supporting Information).  This analytical description is exact in the case of infinitely ``soft'' learning and infinite number of cells.  It is based on the assumption that at each training iteration the distribution of cells is changed by infinitely small amount, correspondingly requiring an infinite number of training iterations to achieve finite changes in cell distribution. Thus the discrete iteration step can be replaced by continuous ``time'' $t$, and the evolution of the number of cells  $n_i$ with a given value $m_i$ of genetic diversity parameter $m_u$,  in the course of training can be described by a differential equation 
$$
{dn_i\over dt}=\lambda_i n_i-{n_i\over N_c}\sum_k \lambda_k n_k 
$$
where $N_c$ is the total number of cells (fixed) and $\lambda_i$ is a ``shaping factor'' which depends on the distributions of the positive and negative training examples $w_\pm(x)$, $m_i$-dependent single cell GFP response functions $z(x; m_i)$ and the corresponding cell survival probabilities $p_\pm(z)$ (see Supporting Information). The solution of this equation approximates an ensemble average solution in the case of finite ``softness'', number of training steps ($N$), and finite number of cells by formally setting $t=N/2$. Figure~\ref{fig.analytics} demonstrates the good agreement between the analytical solutions for the ensemble- averaged $m_u$ distribution, population average GFP response, and the ensemble average classifier performance and the simulation results for ``soft'' learning with $\gamma=1$. Note that the performance figures calculated analytically are consistently higher than the ensemble average performance figures calculated from the simulations~(Fig.~\ref{fig.analytics}C,~F). This is largely due to the fact that the analytical approximation is based on asymptotic analytical solution which ignores the stochastic effects due to the finite number of iterations.  

This analytical theory allows to quickly estimate the parameters for optimal training of the distributed cell population based classifier for a given classification problem. In comparison the stochastic simulations necessary to estimate the same training parameters are significantly more computationally expensive. If necessary the results of the analytical approximation can be confirmed by stochastic simulations.

Another important result that follows from the analytical approximation is that early stopping must be always used in order to maximize the classifier performance, even in the case of infinitely ``soft'' learning, since after large number of training iterations ($N \rightarrow \infty$) only the cells with the maximum $\lambda_i$ survive. In most practical cases it means that the cells with only one particular value of $m_i$ survive thus generally leading to poor classifier performance.


\subsection{Summary and Concluding Remarks}
\begin{figure}[ht] 

  \includegraphics[width=\linewidth]{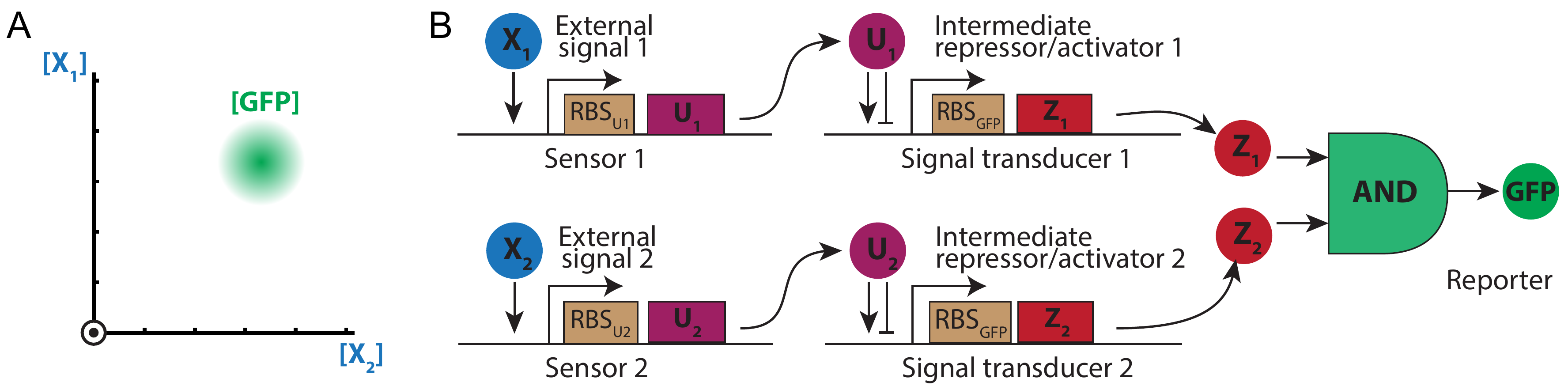}
 
  \caption{ (A) The desired bell-shaped response of the multidimensional classification circuit. (B) The genetic circuit proposed for use in a distributed genetic classifier with multiple inputs per cell. Independent sensing and response functionalities are combined using an appropriate biological AND gate. The resulting response function of the entire two-promoter circuit is bell-shaped with respect to the inputs $X_1$ and $X_2$ for the relevant choice of parameters.}
  \label{fig3ad}
\end{figure} 

In this paper we introduced a conceptual design of a distributed classifier in the form of a population of microorganisms containing a strategically constructed library of sensory gene circuits. We described an algorithm of training such a classifier by pruning the master population after iterative presentation of known positive and negative input-output examples. We characterized both numerically and analytically the performance of the proposed classifier based on a particular sensory gene circuit that is induced by the external chemical input within a certain range of concentrations and repressed outside of it. A library containing a broad distribution of such circuits with different sensitivities to the chemical input in individual cells within the master population can be constructed by randomizing the control sequences within the input genetic element of the sensing circuit. We demonstrated that after appropriate training the distributed classifier can achieve nearly-optimal performance in solving the task of discriminating nonseparable input data sets. 

In this theoretical study we have not addressed in detail several important issues that are likely to arise in the experimental implementation of the distributed classifier. One such issue is  the specific procedure to retain the ``good'' cells and discard the ``bad'' ones. The most straightforward way to do this is to use a fluorescence-activated cell sorting (FACS). A potentially more interesting approach allowing for autonomous and adaptable training could be to engineer an additional circuit controlling cell growth based on the output signal, however we did not explore this possibility here. Further, we assumed that the promoter strength parameters are initially distributed uniformly over a broad range, however experimentally the distribution will likely be non-uniform. This may limit the ability to train the classifier to an arbitrary classification task, although the training algorithm can be adapted to minimize the performance degradation. Finally, we did not take into consideration the growth and division of cells that can unintendedly change the distribution of cells within the trained population if cells with different sensory circuit parameters differ in their doubling rates, for example due to respective differences in metabolic burden exerted by these circuits. 

The particular design of a distributed classifier presented here is suitable for classification of a scalar input (single chemical inducer $X$) only. Many real-world classification problems involve multidimensional inputs. The one-dimensional classifier described in this paper can be generalized to solve multidimensional problems. In the simplest case of multidimensional classification where each dimension can be classified independently, the problem can be trivially solved by a cell population classifier consisting of a mixture of cells capable of responding individually to just one input using the circuit described in Fig.~\ref{fig1ad}. However, such approach would fail for more complex classification problems when a multidimensional distribution of positive or negative outcomes is not the direct product of corresponding one-dimensional distributions. These more complex problems can be solved using a classifier built with the cells endowed with circuits that are sensitive to multiple inputs. An example of such circuit design for a two-dimensional input is shown in Fig.~\ref{fig3ad}. In this design the input signals are sensed separately by the corresponding two-stage modules similar to described above. The outputs of these two modules are then combined by a genetic AND gate. A  number of circuits performing logical operations including AND have been developed and characterized recently~\cite{wang2011engineering, moon2012genetic, shis2013library}.  Similarly to the one-dimensional circuit the output of this circuit is a two-dimensional bell-shaped function of the two inputs (Fig.~\ref{fig3ad}A). The parameters of both sensory circuits can be randomized as before. These multidimensional classifiers can be trained and the performance can be characterized in the same way as in the case of the one-dimensional classifier. 

From synthetic biology perspective, our modeling study opens an intriguing possibility to engineer genetically diversified microbial populations to solve complex classification tasks which are difficult or impossible to solve within a single microbial cell. Such approach is known in machine learning theory where it was proven that consensus classifiers made of appropriately combined weak classifiers can achieve high performance \cite{freund1999short,koltchinskii2002empirical,zhang2012coherence}. On a more general level, our findings suggest that perhaps the genetic and phenotypic diversity found in many natural populations \cite{nevo2001evolution,avery2006microbial}, along with other potential survival benefits \cite{fraser2009chance} can play an important role in forming a decentralized ``social intelligence'' \cite{jacob2004bacterial,couzin2009collective} capable of solving complex computational tasks in a noisy and unpredictable environment.

\begin{acknowledgement}
Author wish to thank R. Kotelnikov and A. Zaikin for useful discussions. This work was partially supported by NIH Grant RO1-GM069811 (LT, JH), the San Diego Center for Systems Biology, NIH Grant P50 GM085764 (AD), and Russian Foundation for Basic Research grant RFBR 13-02-00918 (OK, MI).
\end{acknowledgement}


\bibliography{ensemble_classifier}

\end{document}